\documentstyle[floats,graphicx,twocolumn,aps]{revtex}

\begin{document}
\draft
\preprint{HEP/123-qed}
\title{Magnetic Penetration Depth and Surface Resistance in 
     YBa$_2$Cu$_3$O$_{7-\delta}$: New Results for Ultra High Purity Crystals}
\author{S. Kamal, Ruixing Liang, A. Hosseini, D.A. Bonn, and W.N. Hardy}
\address{Department of Physics, University of British Columbia, Vancouver,
         B.C., Canada  V6T 1Z1}

\date{\today}
\maketitle
\begin{abstract}

We have succeeded in growing very high purity (99.995\%)
$YBa_2Cu_3O_{7-\delta}$ crystals
in $BaZrO_3$ crucibles  
and have measured $\Delta\lambda(T)$ and $R_s(T)$ at
1~GHz in crystals with various oxygen treatments.
For an oxygen vacancy level of $\delta$=0.007, 
$\Delta\lambda$ and $R_s$ essentially
reproduce our previous results and show no sign of
the existence of the two order parameter components as 
recently reported by Srikanth {\it et al.} on $BaZrO_3$-grown crystals.
For other oxygen concentrations, we have in some cases observed 
deviations from the linear low T dependence of $\Delta\lambda$,
but never any sign of a second transition.
\end{abstract}

\narrowtext

Measurements of the electrodynamics of high temperature
 superconductors $(HiT_c)$ have played a crucial role in understanding
 the physics of these materials.
The temperature dependence of magnetic penetration depth $\lambda(T)$
and microwave 
 surface
 resistance $R_s(T)$
 gives information about the nature of quasiparticle
 excitations, their dynamics and, indirectly, information on the
 structure of the gap function.
However, many early attempts at measuring these quantities led to
 misleading conclusions, partly because of problems with sample quality.
The first concern is purity:
 impurities can be introduced into the $HiT_c$ 
material either from the starting chemicals or from the crucible
 during crystal growth.
The second concern is the quality of the surface.
Since electrodynamic measurements involve probe currents that
 flow only within about a thousand angstroms of the surface of the
 crystal, it is natural to raise this concern, and one must 
 distinguish between
 measurements probing the bulk, such as specific heat or thermal 
 conductivity, and those probing the surface such as microwave 
 or infra-red. 

The linear temperature dependence of $\lambda(T)$
was first observed
 by Hardy {\it et al.} for $YBa_2Cu_3O_{7-\delta}$ crystals
 grown in yttria stabilized
 zirconia ($YSZ$) crucibles (purity $\simeq$ 99.9\%) \cite{hardy}.
They found that $\Delta\lambda(T)=\lambda(T)-\lambda(1.2K)$
below 20 K is largely linear
 with a slight, sample-dependent
 curvature below 4 or 5 K. 
Furthermore, studies of deliberate cation substitutions revealed that the
 penetration depth is very sensitive to certain types of impurities:
for example, 0.3\% Zn is enough to change the low temperature behaviour from
 linear to quadratic. Other types of crystal defects might have a similar
 effect; in particular, most films exhibit a T$^2$ behaviour.
It was therefore reasonable to
 believe that the observed 
sample variation in $YSZ$-grown pure crystals was due to the presence of 
impurities or other crystal imperfections.
 
Similar conclusions were drawn concerning the surface resistance of $YBa_2Cu_3O_{7-\delta}$
single crystals.
Bonn {\it et al.} observed a peak in $R_s(T)$ of
$YSZ$-grown crystals below $T_c$ which was attributed to a rapid increase
in quasiparticle scattering time in the superconducting state\cite{bonn}. 
However, the magnitude of the increase could be limited by
deliberately introducing impurities: 
0.3\% Zn and 0.7\% Ni were
shown to be enough to completely suppress the peak.
This left open the question of whether or not the scattering time at low 
 temperatures in $YSZ$-grown
crystals is limited by the residual 0.1\% impurities,
or by some other mechanism.
Like $\lambda(T)$, 
$R_s(T)$ also exhibits considerable sample dependence below 4~K. The
magnitude of the residual $R_s(T)$ at 1.2~K varies considerably
and the temperature
dependence of $R_s(T)$ at low T varies from linear to quadratic in T
\cite{bonn2}.
These all point to the fact that the presence of residual impurities and
crystal defects prevents us from observing some important features of the
intrinsic behaviour of $YBa_2Cu_3O_{7-\delta}$.

A breakthrough in quality of 
 $YBa_2Cu_3O_{7-\delta}$ crystals has
 been made through the use of $BaZrO_3$ crucibles instead of $YSZ$.
 Unlike $YSZ$, the $BaZrO_3$ crucibles are essentially inert and do not 
add measurable impurities to the melt during the growth process. This results
 in crystals with at least one order of magnitude increase in
 purity as well as higher crystallinity.
 Erb {\it et al.} were the first to grow such crystals\cite{erb1} and
 Srikanth {\it et al.} have performed microwave measurements on them\cite{srikanth}.
Recently Ruixing Liang in our group at UBC has succeeded in fabricating
$BaZrO_3$ crucibles and growing
high purity crystals in them. In this paper we present the results of our first
 series of measurements of the microwave surface impedance of this new 
 generation of crystals.

The $YBa_2Cu_3O_{7-\delta}$ crystals are grown by a flux-growth technique
 in $BaZrO_3$ crucibles.
Details of the fabrication of the crucibles and the 
crystal growth are given elsewhere\cite{ruixing}.
The crystals have not only high chemical purity (99.99-99.995\%),
but also a high degree of crystalline perfection 
 as measured by the width of the
 (006) rocking curve, FWHM=0.007$^\circ$ (including 0.003$^\circ$ instrumental resolution), a factor of 3 better than the $YSZ$-grown crystals.
The surface impedance measurements are performed using a
superconducting loop gap resonator
 operating at 1.1 GHz.
The sample is positioned inside the loop such that the RF field is 
 applied parallel to the ab-plane of the crystal. 
This way the currents 
 flow primarily in the ab-plane, with a small contribution 
 from the c-axis currents. In previous studies, we have separated out the
c-axis contribution to $\lambda(T)$\cite{hardy2} and
 most recently have done so for $R_s(T)$ \cite{ahmad}. However, in this
 paper we ignore c-axis contributions, which introduce errors of less
 than 5\% to the results and will not affect qualitative features
 of the temperature dependencies. 

\begin{figure}
\centering
\includegraphics[keepaspectratio,width=3.2in]{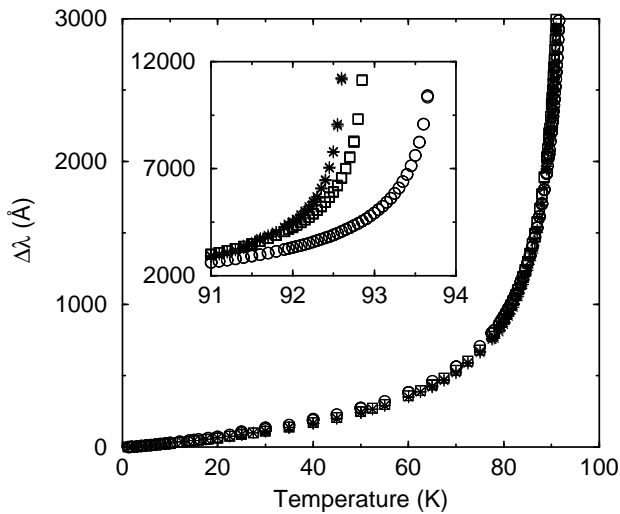}

\caption{$\Delta\lambda(T)$ vs $T$ for twinned crystals of $YBa_2Cu_3O_{7-\delta}$ grown in 
       $BaZrO_3$ crucibles with various annealing temperatures, 500 $^\circ$C
       (circle), 475 $^\circ$C (square) and 450 $^\circ$C (star). Inset shows
       the variation of T$_c$ with annealing temperature.}
\label{fig:dl-anneal1}
\end{figure}

In an attempt to determine optimal conditions for oxygen doping,
we have measured $\Delta\lambda(T)$ for three crystals
annealed in flowing oxygen at temperatures
ranging from 450 to 500 $^\circ$C. 
The areas of the crystals vary between 0.5 to 4 $mm^2$ with thicknesses of
25 to 55 microns. 
The data for all three samples shown in Fig.~\ref{fig:dl-anneal1}
are similar, 
with no indication of the second order parameter component
 reported by Srikanth
 {\it et al.}\cite{srikanth}. 
The inset shows the variation of $T_c$ with annealing temperature, 
with the highest T$_c$ of 93.7 K achieved by annealing  
at 500 $^\circ$C.

\begin{figure}
\centering
\includegraphics[keepaspectratio,width=3.2in]{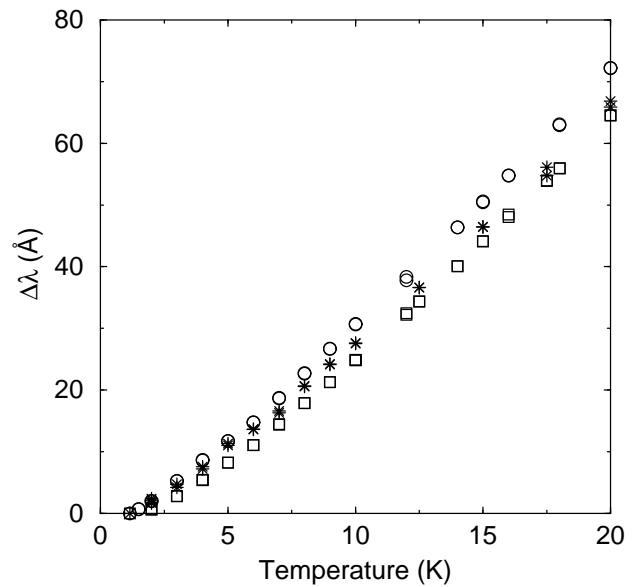}

\caption{Same data as in Fig. \ref{fig:dl-anneal1} shown below 20 K.}
\label{fig:dl-anneal2}
\end{figure}

The same data, shown in detail below 20 K
in Fig. \ref{fig:dl-anneal2}, exhibits curvature which differs
from the largely linear dependencies observed for $YSZ$-grown
crystals.
A likely explanation is that the chain oxygen vacancies 
in the higher purity $BaZrO_3$-grown crystals have a tendency 
to cluster.
Erb {\it et al.}\cite{erb2} have
 proposed that ``fishtail''-shaped magnetization loops observed in their
optimally doped crystals are in fact due to pinning by the vacancy clusters.
From the point of view of microwave measurements,
these clusters could act as electronic scattering
 centers, thus moving $\Delta\lambda(T)$ towards the quadratic temperature
 dependence observed in Zn-doped samples \cite{bonn}.

\begin{figure}

\centering
\includegraphics[keepaspectratio,width=3.2in]{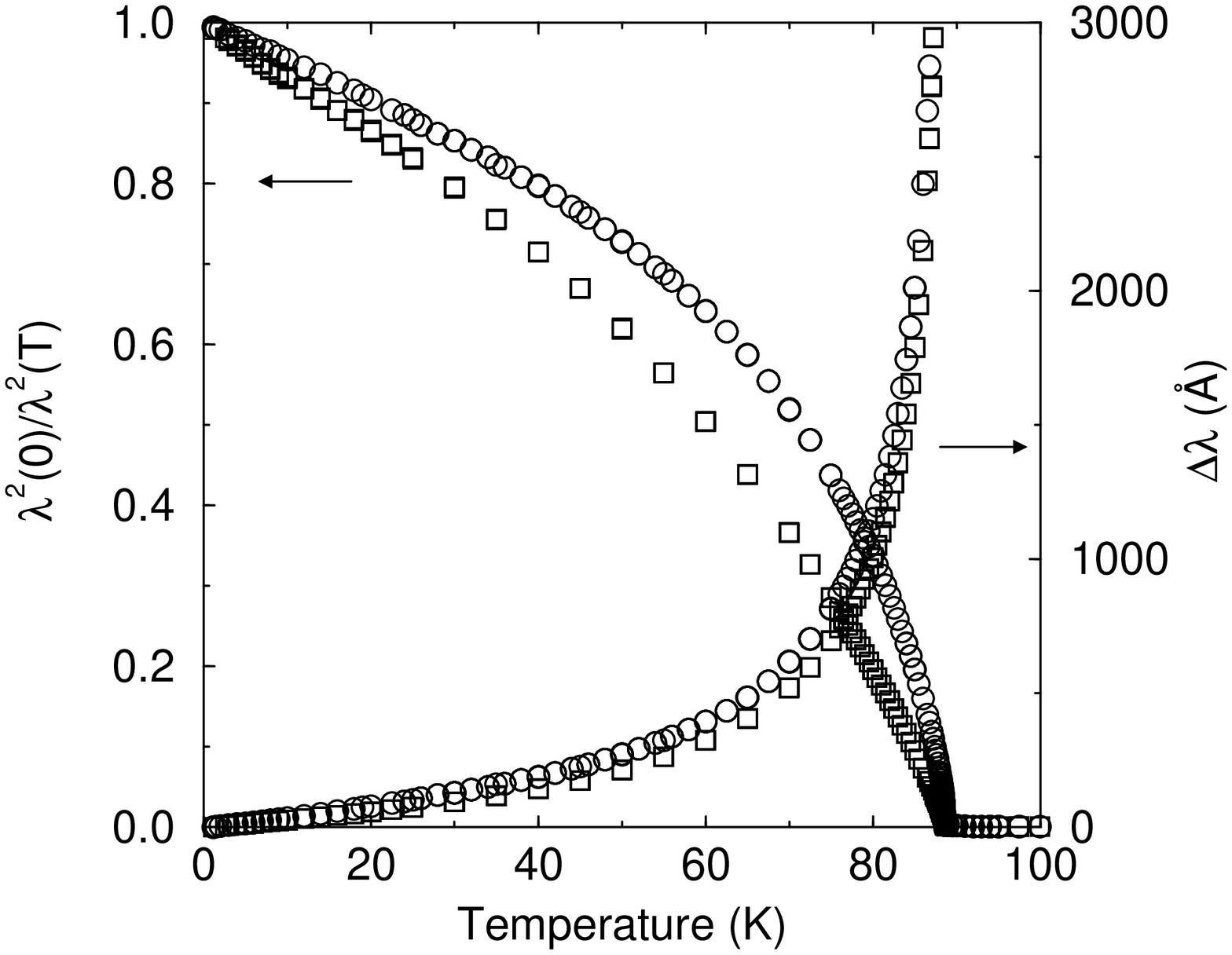}

\caption{$\Delta\lambda(T)$ vs $T$ (right axis) and superfluid fraction
       $\lambda^2(0)/\lambda^2(T)$ 
        vs $T$ (left axis) for a detwinned crystal of 
       ${\rm YBa_2Cu_3O_{6.993}}$, for a-(circle) and b-(square)
        directions.}
\label{fig:dl-overdoped1}

\vskip 1pc
\includegraphics[keepaspectratio,width=3.2in]{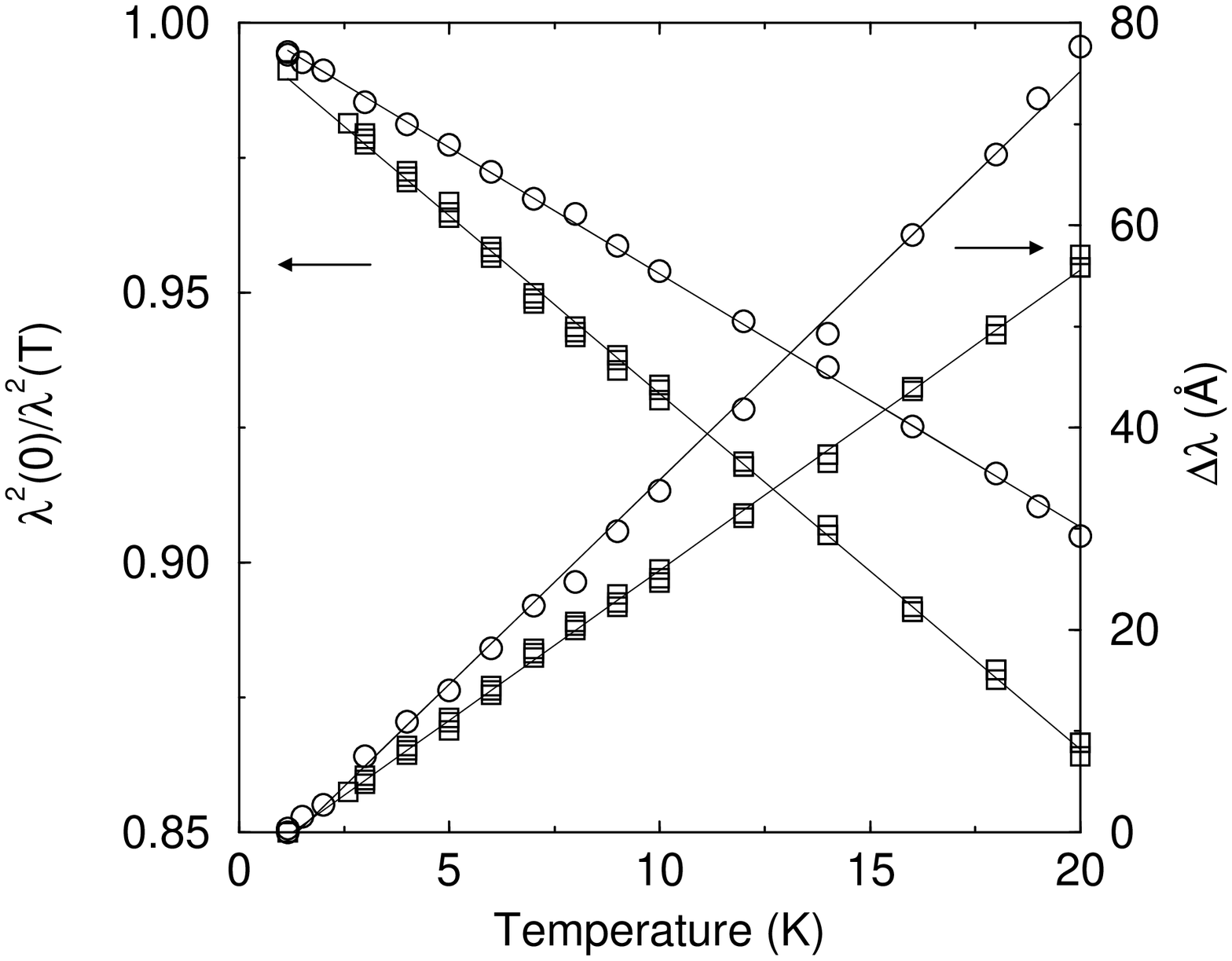}

\caption{Same as figure \ref{fig:dl-overdoped1} but shown for below 20 K.
 All solid
       lines are linear fits to the data.}
\label{fig:dl-overdoped2}
\end{figure}

One obvious way to avoid  clustering is to dope the crystals as close as 
possible to O$_7$, where it has been
shown that the magnetization ``fishtail''
disappears \cite{erb3}.
Figures \ref{fig:dl-overdoped1} and \ref{fig:dl-overdoped2} show the
 results of this approach.
The crystals were detwinned and then annealed for 50 days, with the
annealing temperature initially set at 450 $^\circ$C and then 
 decreased in several steps, the last one being 350 $^\circ$C which
 corresponds to O$_{6.993}$. 
Figure \ref{fig:dl-overdoped1} shows $\Delta\lambda(T)$ and
 the superfluid fraction $\lambda^2(0)/\lambda^2(T)$ for the a- 
and b-directions over the whole temperature range below $T_c \simeq 88.7$ K.
 The $\Delta\lambda$'s are similar to those shown in
 figure 1 and do not differ substantially from data on
 crystals grown in $YSZ$ crucibles. 
As yet, the zero temperature values of penetration depth,
 $\lambda(0)$, of these crystals are not known. 
We have used values $\lambda_a(0)=1600$ \AA \mbox{} and $\lambda_b(0)=800$~\AA \mbox{}
 which are inferred from $\mu$SR measurements for overdoped $YBa_2Cu_3O_{7-\delta}$
 crystals\cite{tallon}. However, the choice of
 $\lambda(0)$ does not affect the qualitative features of the superfluid
 density, namely no signature of a second order
 parameter component developing below T$_c$ as reported by
 Srikanth {\it et al.}, and non-mean field
 behaviour near T$_c$.

Figure 4 shows the detailed behaviour of the data below 20 K, revealing
a slight curvature in $\Delta\lambda$ in the a-direction and
very linear temperature dependence in the b-direction.
Linear fits to $\Delta\lambda(T)$ give slopes of 4.0 \AA/K and 
 3.0 \AA/K for the a- and b-directions respectively, which are very
 similar to the slopes of 4.0 \AA/K and 3.2 \AA/K observed
 for overdoped $YBa_2Cu_3O_{7-\delta}$ crystals grown in $YSZ$ crucibles.
Power law fits to $\lambda^2(0)/\lambda^2(T)$ below 20 K give exponents of 1.06 and 0.94 
for the a- and b-directions respectively, very close to linear.

We call attention to the fact that 
the linear temperature dependence persists down to the 1.15~K base 
temperature, whereas for $YSZ$-grown crystals
we typically observe a cross-over towards higher power laws below 4 or 5 K.
This supports the conjecture that the curvature observed in $YSZ$-grown crystals
is due to the presence of the $\sim$0.1\% impurities:
the new crystals have more than an order of magnitude
 higher purity and correspondingly little curvature.
Kosztin and Leggett\cite{kosztin} have predicted that non-local
 effects can result in deviations from the expected linear temperature
 dependencies even for a pure d-wave superconductor.
However, as they have noted, non-local effects for the ab plane
 penetration depth are mainly important in the geometry where the
 magnetic field is applied parallel to the c-axis. 
In the measurements reported here, the RF field is 
applied parallel to the ab-plane and non-local effects should
 be negligible.

\begin{figure}

\centering
\includegraphics[keepaspectratio,width=3.3in]{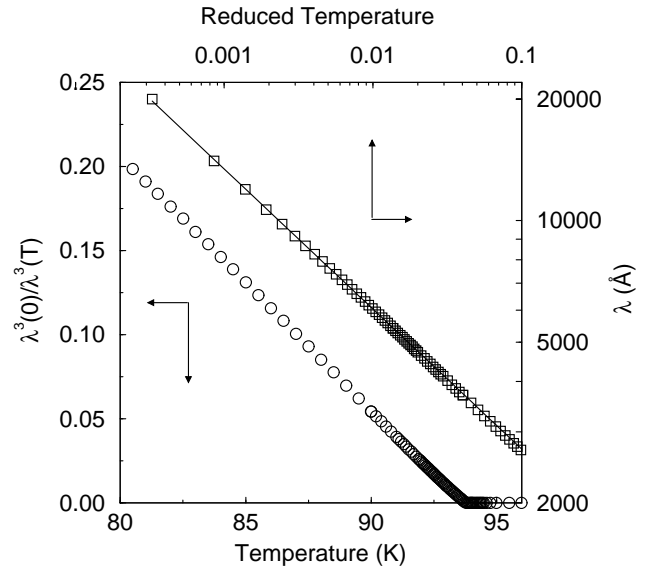}

\caption{3DXY critical behaviour of the superfluid density in
        the superconducting state of a
	sample annealed to $YBa_2Cu_3O_{6.92}$. The circles show
        $\lambda^3(0)/\lambda^3(T)$ vs $T$ (left and bottom axes) and the squares 
        show $\lambda(t)$ vs reduced temperature $t=1-T/T_c$ on
        a log-log scale (right and top axes).}
\label{fig:3dxy}
\end{figure}

Previous measurements of $\lambda(T)$ in our laboratory
  have shown non-mean field behaviour close
 to T$_c$. Kamal {\it et al.} observed that in $YSZ$-grown crystals, the
 superfluid density shows the critical behaviour of the 3DXY
 universality class
 and that, surprisingly, the critical region is as wide as 10 K\cite{saeid}.
 In figure \ref{fig:3dxy} we show $\lambda^3(0)/\lambda^3(T)$ (circles) for
 a twinned $YBa_2Cu_3O_{7-\delta}$ single crystal
 grown under optimal doping conditions in a $BaZrO_3$ crucible.
The crystal was annealed at 500 $^\circ$C to an oxygen content of
$O_{6.92}$, has $T_c\simeq93.78$ K and 
 most importantly, has a very sharp transition of less than 0.25 K wide.
For $\lambda(0)$ we have chosen 1400 \AA, the same value used 
for twinned, $YSZ$-grown crystals, but again the results are not very sensitive
 to this value.
As seen in the figure, this crystal also shows 3DXY critical fluctuations
over a fairly wide temperature range, $\sim$10 K, 
 very similar to $YSZ$-grown crystals.
The squares show a log-log plot of $\lambda$ as a function of
 reduced temperature, $t=1-T/T_c$, over almost 3 decades. The solid line
 is a fit to a power law $\lambda(t)=\lambda_\perp t^{-y}$ with
  $y=0.34\pm.01$, where the
 error corresponds to assuming $\pm$200 \AA \mbox{} error in $\lambda(0)$.

The loop gap resonator and sample holder used in these measurements were
 designed mainly for precision measurements of $\lambda(T)$.
However, we have recently succeeded in making simultaneous 
measurements of surface
 resistance using this resonator, thanks to recent improvements in the
 unloaded Q (Q$_0\simeq 4\times 10^6$) 
and
 to the use of time domain techniques.
In our surface resistance measurements we are usually able to
 withdraw the sample from the resonator in order to find the unloaded Q.
This is not possible in the present configuration and we can only measure 
 the change of surface resistance, $\Delta R_s(T)=R_s(T)-R_s(1.15 K)$. 
However, from other measurements on comparable crystals we know that
 the residual $R_s$ is very low, probably less than 0.5 $\mu\Omega$.

\begin{figure}[t]

\centering
\includegraphics[keepaspectratio,width=3.3in]{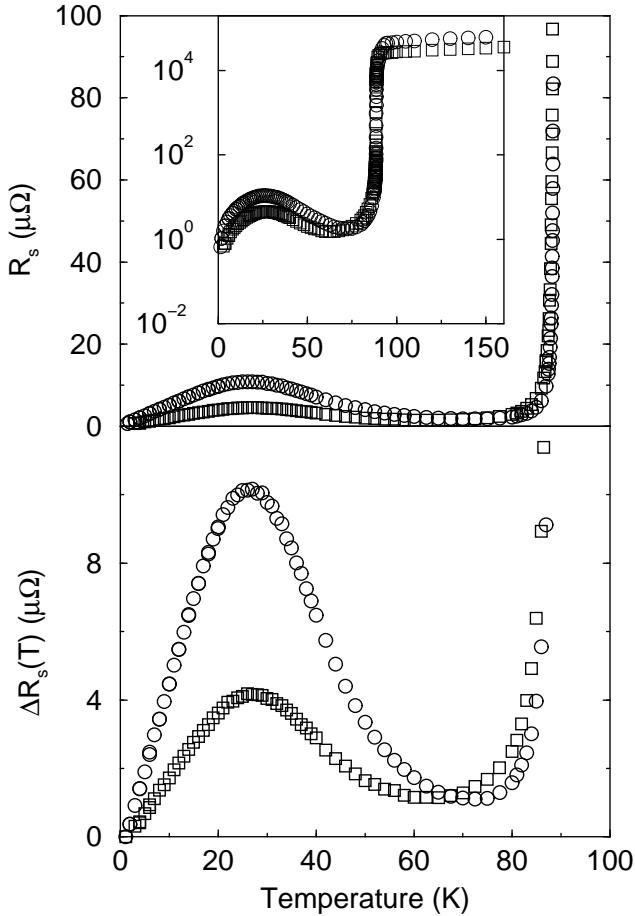}

\caption{Surface resistance vs $T$ for a detwinned crystal of 
       ${\rm YBa_2Cu_3O_{6.993}}$, for a-(circle) and b-(square) directions.}
\label{fig:Rs}
\end{figure}

Figure \ref{fig:Rs} shows R$_s$(T) for the a- and b-directions.
The main features are similar to our previous results on
 crystals grown in $YSZ$ crucibles, where the peak at 26 K is
 attributed to a rapid rise in the  quasiparticle scattering time below
 T$_c$. In particular, there is no indication whatsoever of a second order
 parameter developing below T$_c$ as reported by
 Srikanth {\it et al.}\cite{srikanth}.
Even with the inclusion of a
somewhat uncertain residual surface resistance, one of the 
striking features of the data is that the average
rise in $R_s$ from its minimum at about 70 K to its maximum at 26 K is
roughly four fold compared to the two fold increase in crystals from $YSZ$
crucibles. We interpret this as indicating that in the new crystals the
quasiparticle scattering time rises to a much higher limiting
value than in the
$YSZ$-grown crystals, a consequence of the higher purity of the $BaZrO_3$-grown
crystals.
Another noteworthy difference is that $R_s(T)$ varies linearly with temperature
all the way down to 1.15~K, with
slopes of 0.52 and 0.21 $\mu\Omega/K$ for a- and b-
directions respectively. 

In summary, we have presented
$\Delta\lambda(T)$ and $R_s(T)$ for very high purity crystals
of $YBa_2Cu_3O_{7-\delta}$ grown 
 in $BaZrO_3$ crucibles.
The results show no evidence for two order parameter components.
 This is consistent with the fact that the specific heat of
 $BaZrO_3$-grown crystals produced by Erb {\it et al}. \cite{erb3}
 exhibits a peak at $T_c$ that
 is identical in size and shape to those seen in high quality $YSZ$-grown
 crystals \cite{ruixing2}.
 The specific heat data on the new crystals also shows no sign of a second
 superconducting phase
 transition. It is difficult to reconcile this {\it bulk} measurement on
 the $BaZrO_3$-crystals with the surface impedance data of Srikanth {\it et al}.
 \cite{srikanth} which shows a rather weak increase in the superfluid
 density near $T_c$ and a new feature at lower temperatures. This disagreement
 with a bulk measurement,
 coupled with the fact that we have observed no sign of a second phase in our
 surface impedance measurements, forces us to conclude that all of the new
 features reported by Srikanth {\it et al.} arise from some sort of
 problem with the surfaces of the crystals that they have been studying.

\acknowledgments
The authors acknowledge helpful discussions with A. Erb and
 R. Hackl and the assistance of P. Dosanjh.
This work was supported by the National Science and
 Engineering Research Council of Canada and the Canadian Institute
 for Advanced Research. DAB gratefully acknowledges support from
 the Sloan Foundation.

\end{document}